\documentclass[%
 reprint,
 amsmath,amssymb,
prb,
]{revtex4-1}

\usepackage{multirow}
\usepackage[normalem]{ulem}

\usepackage{longtable}

\usepackage{threeparttable}

\usepackage{graphicx}
\usepackage{dcolumn}
\usepackage{bm}

\usepackage[english]{babel}
\usepackage[utf8]{inputenc}
\usepackage{amsmath}
\usepackage{amsfonts}

\usepackage{natbib}

\usepackage{graphicx}
\usepackage[colorinlistoftodos]{todonotes}
\usepackage{lmodern}

\usepackage{epstopdf}
\usepackage{hyperref}

\usepackage{braket}

\def\beq{\begin{equation}}
\def\eeq{\end{equation}}

\def\bs{\begin{split}}
\def\es{\end{split}}

\def\bea{\begin{eqnarray}}
\def\eea{\end{eqnarray}}

\date{\today}

\begin{document}

\preprint{APS/123-QED}

\title{Understanding molecular representations in machine learning: 
The role of uniqueness and target similarity}


\author{Bing Huang} 
\author{O. Anatole von Lilienfeld}
\email{anatole.vonlilienfeld@unibas.ch}
\affiliation{%
Institute of Physical Chemistry and National Center for Computational Design and Discovery of Novel Materials (MARVEL),
Department of Chemistry, University of Basel, Klingelbergstrasse 80, 4056 Basel, Switzerland}%

\date{\today}

\begin{abstract}
The predictive accuracy of Machine Learning (ML) models of molecular properties depends on the choice of the molecular representation.
Based on the postulates of quantum mechanics, 
we introduce a hierarchy of representations which meet uniqueness and target similarity criteria.
To systematically control target similarity, we rely on interatomic many body expansions,
as implemented in universal force-fields, including
{\underline B}onding, {\underline A}ngular, and higher order terms (BA). 
Addition of higher order contributions systematically increases similarity to the true potential energy 
{\em and} predictive accuracy of the resulting ML models. 
We report numerical evidence for the performance of BAML models trained on molecular properties pre-calculated 
at electron-correlated and  density functional theory level of theory for thousands of small organic molecules. 
Properties studied include 
enthalpies and free energies of atomization, heatcapacity, zero-point vibrational energies,
dipole-moment, polarizability, HOMO/LUMO energies and gap, ionization potential, electron affinity, and electronic excitations. 
After training, BAML predicts energies or electronic properties of
out-of-sample molecules with unprecedented accuracy and speed. 

\end{abstract}

\pacs{Valid PACS appear here}
\maketitle


Reasonable predictions of ground-state properties of molecules require computationally demanding calculations of approximated expectation values of the corresponding operators. \cite{szabo1989modern} 
Alternatively, Kernel-Ridge-Regression (KRR) based machine learning (ML) models \cite{Muller1996}
can also {\em infer} the observable in terms of a linear expansion in chemical compound space~\cite{CCS,anatole_qc2013}.
More specifically, an observable can be estimated using 
$O^{\mathrm{inf}}({\bf M}) = \sum_i^N \alpha_i k(d(\mathbf{M},\mathbf{M}_i))$, where $k$ is the kernel function (we use Laplacian kernels with training set dependent width), $d$ is a metric (we use the $L_1$-norm), 
and ${\bf M}$ is the molecular representation~\cite{ML0,njp}. 
The sum runs over all reference molecules $i$ used for training to obtain regression weights $\{\alpha_i\}$. 
Once trained, the advantage of such ML methods consists of (i) their computational efficiency (multiple orders of speed-up with respect to conventional quantum chemistry) and (ii) their accuracy can systematically be converged to complete basis set limit through addition of sufficient training instances. 
Their drawback is that they (a) are incapable of extrapolation by construction, 
and (b) require substantial training data before reaching satisfying predictive power for out-of-sample molecules. 
In practice, addressing (a) is less important since one typically knows beforehand which ranges of interatomic distances and chemical compositions are relevant to the chemical problem at hand:  It is straightforward to define the appropriate domain of applicability for the application of supervised ML models in chemistry.
In recent years, much work has been devoted to tackle the latter drawback through the discovery and development of improved representations ${\bf M}$ \cite{Behlers_descriptor,Bartok_Csanyi_descriptor,Muller_Gross_crystal,OAvL_FRD,Scheffler_descriptor_design,qc_Felix_crystal,   Ramprasad_descriptor}.

\begin{figure} 
\centering
\includegraphics[scale=0.17]{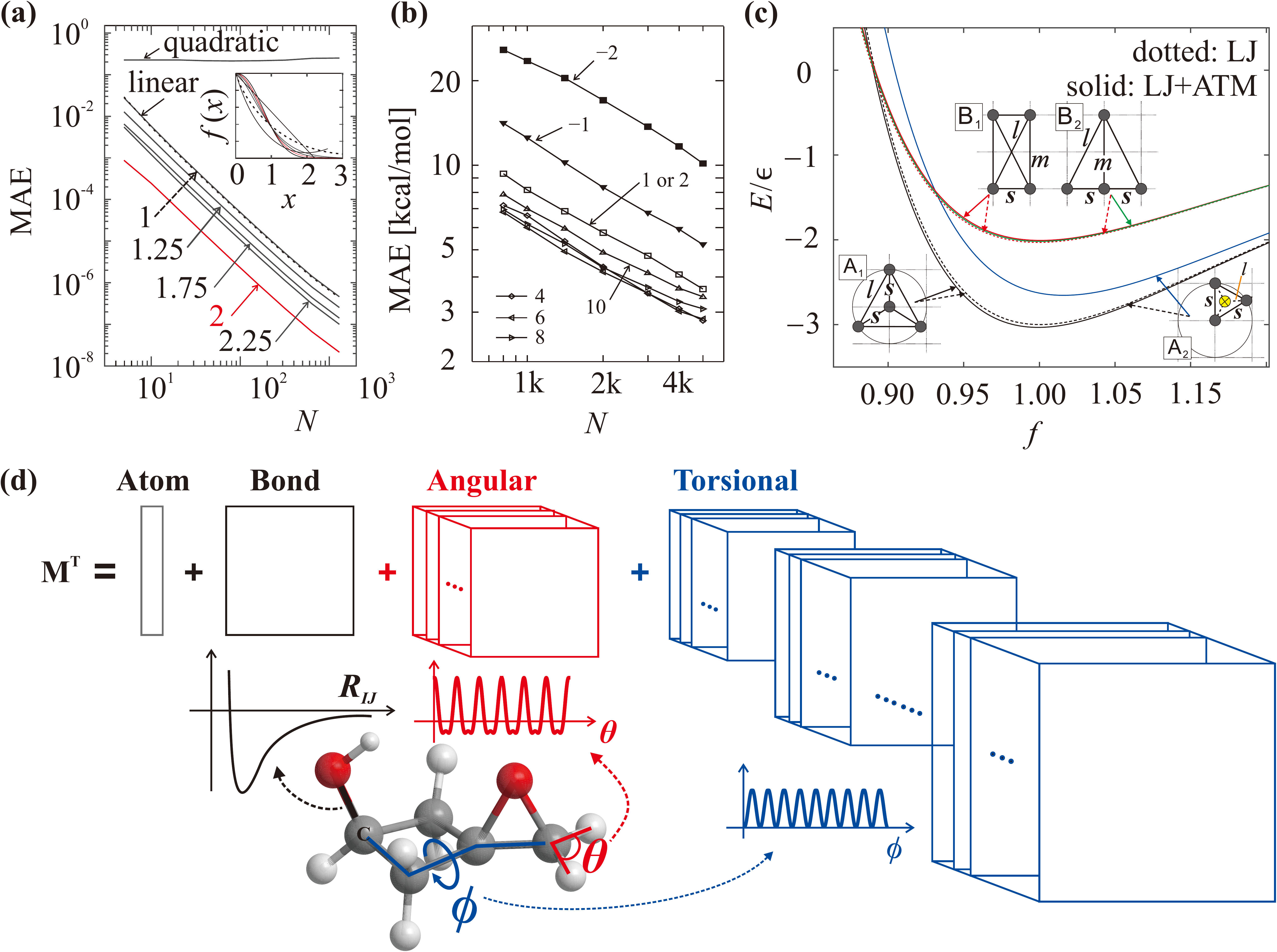}
\caption{\label{fig:Representation} 
Target similarity determines offset $a$ in ML model learning curves $\log({\rm Error}) = a - b \log(N)$.
Panels (a) and (b) illustrate learning curves for ML models obtained for representations of varying target similarity applied to (a) modeling a 1-D Gaussian target function or (b) enthalpy of atomization for QM7b dataset \cite{njp}. 
Lines in (a) correspond to models resulting from linear, quadratic, and various exponential ($e^{-x^n}$ with $n$ = $\{$1, 1.25, 1.75, 2 and 2.25$\}$) representations. The inset shows the target function (red) as well as the representations.
Learning curves in (b) correspond to models resulting from Coulomb matrices with varying definition of off-diagonal elements, $Z_IZ_J/R_{IJ}^n$, where $n$ is specified in the figure. 
(c) Illustration that 3-body interactions are crucial for distinguishing two pairs of homometric Ar$_4$ clusters: pair A (A$_1$ and A$_2$, where $l=\sqrt{3}s$) and pair B (B$_1$ and B$_2$, where $m=2s, l=\sqrt{5}s$). 
Horizontal axis label $f$ scales $s$, where $f=1$ corresponds to the choice $s=3.82$ $\mathrm{\AA}$. 
LJ and ATM correspond to Lennard-Jones and Axilrod-Teller-Muto~\cite{atm,atm2} potentials, respectively.
(d) Illustration of the universal force-field~\cite{uff} based construction of the BA representation. 
}
\end{figure}


Often, new descriptors are introduced based on {\em ad hoc}/trial-and-error reasoning, 
such as using features which characterize a system or the property of interest, which meet known invariances, which simply add ``more  physics'', or which simply encode anything which could possibly be related to the system and property of interest. 
The use of new representations is justified {\em a posteriori}: They yield ML models with good performance. Severe survivorship bias, however, might occur: Alternative representations which could have been obtained from the same set of guidelines, and which would result in poor ML model performance, are ommitted. 
While such practice is not wrong it is questionable: Arbitrarily many different representations could have been 
designed using the exact same guidelines. 
To the best of our knowledge, there is no general specific and rigorous procedure for systematically optimizing molecular rexpresentations for converged and robust ML model performance {\em without} drawing heavy numerical experimentation and conclusions {\em a posteriori}. 
In this paper, we rely on the postulates of quantum mechanics to identify a hierarchy of representations 
which enable the generation of ML models with systematically increased predictive accuracy. 
We approximate the hierarchy using universal force-field (UFF) parameters~\cite{uff}.
To support our findings, we present numerical evidence for multiple ML models 
using the same set of quantum chemical ground-state properties and structures, previously calculated for
thousands of organic molecules \cite{gdb9} 

For large $N$, errors of ML models have been observed to decay as inverse roots of $N$ \cite{Muller1996},
implying a linear relationship, $\log({\rm Error}) = a - b \log(N)$.
Therefore, the best representation must
(i) minimize the off-set $a$ and (ii) preserve the linearity in the second term while maximizing its pre-factor $b$.
According to the first postulate of quantum mechanics any system is represented by its wavefunction $\Psi$ which results from applying the variational principle to the expectation value of the Hamiltonian operator. 
As such, there is a one-to-one relationship between Hamiltonian and $\Psi$. 
While some representations have been introduced in order to mimic $\Psi$ (or its corresponding electron density~\cite{hk}) \cite{SOAP,mallathirsch2016}.

Unfortunately, many observables are extremely sensitive to minute changes in $\Psi$, as such we prefer 
to focus directly on the system's Hamiltonian (and the potential energy surface it defines) defined in chemical compound space~\cite{anatole_qc2013}.  
We have realized that representations based on increasingly more accurate approximations to the potential energy surface afford increasingly more accurate KRR ML models. 
In other words, as tbetween representation and potential energy (target similarity) increases
the off-set $a$ in the linear $\log-\log$ learning curve decreases.
Furthermore, $b$ appears to be a global constant, independent of the representation's target similarity, 
as long as the crucial~\cite{OAvL_FRD} uniqueness criterion is met.

First, we exemplify the importance of target similarity for a mock supervised learning task: Modeling a 1D Gaussian function (Inset Fig.~\ref{fig:Representation}(a)). 
As representations ${\bf M}$ we use  linear, quadratic, and exponentially decaying functions with varying exponent of $x$.
Learning curves of resulting ML models (Fig.~\ref{fig:Representation}(a)) indicate systematic improvement as the target similarity, i.e.~similarity of representation to Gaussian function, increases. 
Note that all learning curves, with the notable exception of the quadratic one, exhibit the same slope $b$ on the log-log plot of the learning curve: They only differ in learning curve off-set $a$ which co-incides with their target similarity.
When using a Gaussian function as a representation, the smallest off-set is observed---as one would expect.
It is easy to see why the error of the ML model using the quadratic function as a representation does not decrease when adding more training data: Its minimum is at $x=2$, 
and in the region $x > 2$ the function turns upward again, preventing a one-to-one map between $x$ and representation. 
In other words, the quadratic function is not monotonic and therefore lacks uniqueness, 
introducing noise in the data which can not converge to zero and which results
in a constant error for large $N$. 
By contrast, all other representations are monotonic and conserve the one-to-one map to $x$. 
As such, they are unique representations and they all reduce the logarithm of the
error in a linear fashion at the same rate as the amount of training data grows. 
While the rate appears to be solely determined by the uniqueness of the representation, 
confirming that uniqueness is a necessary condition for functional descriptors~\cite{anatole_qc2013}, 
the off-set $a$ of the resulting learning curve appears to be solely determined by target similarity. 

To see if our line of reasoning also holds for real molecules, 
we have investigated the performance of ML models for predicting atomization energies of organic molecules using a set of unique representations with differing target similarity. 
More specifically, we calculated learning curves for ML models resulting from 
atom adjacency matrices derived from the Coulomb matrix~\cite{ML0}, with off-diagonal elements $M_{IJ} = Z_IZ_J/R^n_{IJ}$.
Where $R_{IJ}$ is the interatomic distance between atoms $I$ and $J$, and the conventional variant (giving rise to the name) is recovered for $n = 1$.
For any non-zero choice of $n$ these matrices encode the complete polyhedron defined in the high-dimensional space spanned by all atoms in the molecules: They uniquely encode the molecule's geometry and composition, thereby ensuring a constant $b$. 
For negative $n$ values, however, this representation becomes an unphysical model of the atomization energy: The magnitude of its off-diagonal elements {\em increases} with interatomic distance. 
Corresponding learning curves shown in Fig.~\ref{fig:Representation}(b)  reflect this fact:
As off-diagnoal elements become increasingly unphysical by dialing in square-root, linear, and quadratic functions in interatomic distance, respectively, the off-set $a$ increases.  
Conversely, matrix representations with off-diagonal elements which follow the Coulomb and higher inverse power laws 
are more physical and exhibit lower off-sets. 
Interestingly, we note the additional improvement as we change from Coulombic $1/R$ to van der Waals $1/R^6$ like power laws. These results suggest that---after scaling---pairwise London dispersion kind of interactions are more similar to molecular atomization energies than simple Coulomb interactions. 
In the following, we dubb the resulting representation the London Matrix (LM).

The bag-of-bond (BoB) representation, a stripped down pair-wise variant of the Coulomb matrix, has resulted in remarkably predictive ML models~\cite{bob}.
Starting with the insights gained from the above, we use the bagging idea as a starting point for the development of our systematically improved representation. 
Unfortunately, when relying on bags of pair-wise interactions as a representation the uniqueness
requirement is violated by arbitrarily many sets of geometries, 
no matter how strong the (effective) target similarity of the employed functional form. 
In Fig.~\ref{fig:Representation}(d), we demonstrate this issue for two pairs of homometric molecules, each with four rare gas atoms, and 
once in a competition of a pyramidal/planar geometry (A), and once for a rectangular/triangular pair (B). 
Both pairs exhibit the exact same list of interatomic distances: 3 $s$/3 $l$ for A, and 2 $s$/2 $m$/2 $l$ for B.
Consequently, when using a pair-wise energy expression (no matter how effective), the predicted curve as a function
of a global scaling factor $f$ will be indistinguishable for both pairs (example shown in Fig.~\ref{fig:Representation}(d) using Lennard-Jones potentials with parameters for Argon). 
Only after the addition of the corresponding three-body van der Waals Axilrod-Teller-Muto~\cite{atm,atm2} contribution, the homometric pairs can be distinguished~\cite{SIDENOTE}. 

Using the insight gained, we have investigated a hierarchy of representations which consists of bags of 
(1) dressed atoms ($\mathbf{M}^{\mathrm{D}}$), 
(2) atoms and bonds ($\mathbf{M}^{\mathrm{B}}$), 
(2) atoms, bonds and angles ($\mathbf{M}^{\mathrm{A}}$), 
and (3) atoms, bonds, angles, and torsions ($\mathbf{M}^{\mathrm{T}}$). 
To indicate the many-body expansion character we dubb these feature vectors 
``BA-representation'' (standing for bags of {\underline B}onds, {\underline A}ngles, Torsions, etc. pp.). 
Here, we have chosen functional forms and parameters for BA-representations that correspond to UFF~\cite{uff}. 
The corresponding terms are illustrated in Fig.~\ref{fig:Representation}(d), 
and correspond to averaged atomic contributions to energies of molecules in training set for atoms,
Morse and Lennard-Jones potentials for covalent and non-covalent intramolecular bonding, respectively, as well as sinusoidal functions for angles and torsions. 
More technical details regarding the training can be found in the supplementary materials. 
While we find that use of the UFF results in consistent and remarkable performance, we note that other force fields could have been used just as well. 
In particular, we also tested  Lennard-Jones and Axilrod-Teller-Muto potentials
in BA based ML models (BAML), and found systematic improvement and similar performance also for these functional forms and parameters. 

\begin{table*}[]
\centering
\caption{Mean absolute errors and root mean square errors in brackets)
for the ML predictions of 14 molecular properties of molecules in QM7b data set~\cite{njp}.
Results from this work (BAML, BoB, BoL, CM, LM) are shown
together with previously published estimation (SOAP \cite{SOAP}, rand CM\cite{njp}) 
for the same dataset. 
Errors are measured on test set of 2200 randomly selected configurations, while the remaining compounds out of QM7b were used for training. 
Labels specify property and level of theory: Atomization energy ($E$), averaged molecular polarizability ($\alpha$), HOMO and LUMO eigenvalues, ionization potential (IP), electron affinity (EA), first excitation energy ($E_{1^{\mathrm{st}}}^{*}$), excitation frequency of maximal absorption ($E^{*}_{\mathrm{max}}$) and the corresponding maximal absorption intensity ($I_{\mathrm{max}}$). 
Expected averaged deviation from experiment is specified in the last column. 
}
\label{qm7b_comparison}
\resizebox{\textwidth}{!}{
\begin{tabular}{lccccccccccc|ccc}

\hline
property                               & SD     & BAML                                                      & BoB                                      & BoL           & CM            & LM              & SOAP \cite{SOAP}                                        & rand CM \cite{njp}                                      & accuracy  \\ \cline{1-10}
                                                                                                                                                                                                                                                                                                                                                                                                
$E$ (PBE0) [kcal/mol]                  & 223.69 & 1.15 (2.54)                                               & 1.84 (4.15)                              & 1.77 (4.07)   & 3.69 (5.77)   &   2.84 (4.94)   & $\boldsymbol{0.92}$ ({\color{red}$\boldsymbol{1.61}$})  & 3.69 (8.30)                                             &  3.46$^\mathrm{a}$, 5.30$^\mathrm{b}$, 2.08-5.07$^\mathrm{c}$ \\
$\alpha$ (PBE0) [$\mathrm{\AA}^3$]     &  1.34  & 0.07 (0.12)                                               & 0.09 (0.13)                              & 0.10 (0.15)   & 0.13 (0.19)   &  0.15 (0.20)    & $\boldsymbol{0.05}$ ({\color{red}$\boldsymbol{0.07}$})  & 0.11 (0.18)                                             &  0.05-0.27$^\mathrm{d}$, 0.04-0.14$^\mathrm{e}$    \\
HOMO (GW) [eV]                         &  0.70  & $\boldsymbol{0.10}$ ({\color{red}$\boldsymbol{0.16}$})    & 0.15 (0.20)                              & 0.15 (0.20)   & 0.22 (0.29)   &  0.20 (0.26)    & 0.12 (0.17)                                             & 0.16 (0.22)                                             &  -    \\ 
LUMO (GW) [eV]                         &  0.48  & $\boldsymbol{0.11}$ ({\color{red}$\boldsymbol{0.16}$})    & 0.16 (0.22)                              & 0.16 (0.22)   & 0.21 (0.27)   &   0.19 (0.25)   & 0.12 (0.17)                                             & 0.13 (0.21)                                             & -   \\ 
IP (ZINDO) [eV]                        &  0.96  & $\boldsymbol{0.15}$ ({\color{red}$\boldsymbol{0.24}$})    & 0.20 (0.28)                              & 0.20 (0.28)   & 0.33 (0.44)   &   0.31 (0.41)   & 0.19 (0.28)                                             & 0.17 (0.26)                                             & 0.20, 0.15$^\mathrm{d}$   \\ 
EA (ZINDO) [eV]                        &  1.41  & $\boldsymbol{0.07}$ ({\color{red}$\boldsymbol{0.12}$})    & 0.17 (0.23)                              & 0.18 (0.24)   & 0.31 (0.40)   &  0.25 (0.33)    & 0.13 (0.18)                                             & 0.11 (0.18)                                             &  0.16$^\mathrm{f}$, 0.11$^\mathrm{d}$     \\ 
$E_{1^{\mathrm{st}}}^{*}$ (ZINDO) [eV] &  1.87  & $\boldsymbol{0.13}$ (0.51)                                & 0.21 ({\color{red}$\boldsymbol{0.30}$})  & 0.22 (0.31)   & 0.42 (0.57)   &   0.35 (0.46)   & 0.18 (0.41)                                             & $\boldsymbol{0.13}$ (0.31)                              & 0.18$^\mathrm{f}$, 0.21$^\mathrm{g}$   \\ 
$E^{*}_{\mathrm{max}}$ (ZINDO) [eV]    &  2.82  & 1.35 (1.98)                                               & 1.40 (1.91)                              & 1.47 (2.02)   & 1.58 (2.05)   &   1.68 (2.20)   & 1.56 (2.16)                                             & $\boldsymbol{1.06}$ ({\color{red}$\boldsymbol{1.76}$})  & -    \\ 
$I_{\mathrm{max}}$ (ZINDO)             &  0.22  & $\boldsymbol{0.07}$ ({\color{red}$\boldsymbol{0.11}$})    & 0.08 (0.12)                              & 0.08 (0.12)   & 0.09 (0.13)   &  0.09 (0.13)    & 0.08 (0.12)                                             & $\boldsymbol{0.07}$ (0.12)                              &  - \\  \hline

\end{tabular} }
  \footnotetext[0]{$^\mathrm{a}$ MAE of formation enthalpy for the G3/99 set  by PBE0 \cite{error_PBE0_1, error_PBE0_2}; $^\mathrm{b}$ MAE of atomization energy (AE) for 6 small molecules \cite{error_PBE0_6mol_1,error_PBE0_6mol_2} by PBE0; $^\mathrm{c}$ MAE of AE from various studies \cite{error_B3LYP_DFT_Guide} by B3LYP; $^\mathrm{d}$ MAE from various studies \cite{error_B3LYP_DFT_Guide} by B3LYP; $^\mathrm{e}$ MAE from various studies by MP2 \cite{error_B3LYP_DFT_Guide}; $^\mathrm{f}$ MAE for the G3/99 set by PBE0 \cite{error_PBE0_1, error_PBE0_2}; $^\mathrm{g}$ MAE for a set of 17 retinal analogues by TD-DFT(PBE0) \cite{error_TDDFT_PBE0}. }  
\end{table*}

\begin{figure}[htp]
  \centering
  \includegraphics[width=0.5\textwidth]{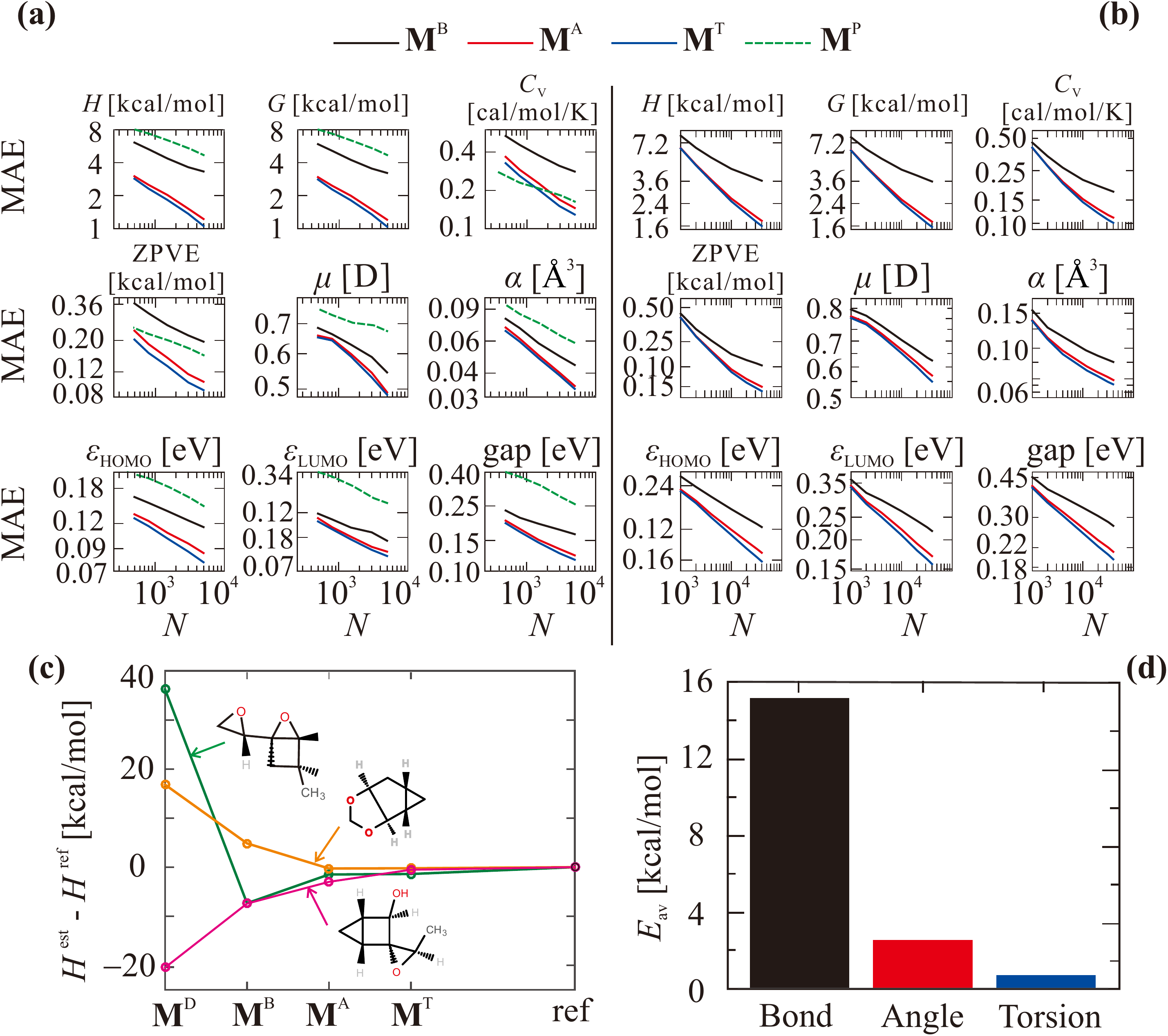} 
  \caption{(a): BAML and polarizability representation based ML learning curves for 9 molecular properties of 6k constitutional isomers of formula C$_7$H$_{10}$O$_2$ \cite{gdb9}. 
  (b) BAML learning curves for 134k QM9 molecules for the same 9 molecular properties \cite{gdb9}. 
  Property models cover $H$, $G$, $C_{\mathrm{V}}$, ZPVE, $\mu$, $\alpha$, $\varepsilon_{\mathrm{HOMO}}$ and  $\varepsilon_{\mathrm{LUMO}}$, i.e.~enthalpy, free energy, heat capacity, zero point vibrational energy, dipole moment, polarizability, HOMO and LUMO energy, respectively. 
  Panel (c) shows convergence of estimated enthalpy values to reference values (all shifted to zero) for three most extreme outlier isomers in C$_7$H$_{10}$O$_2$ set. 
  Panel (d) is the averaged ``contribution'' of each order type in a many-body potential, i.e., the bond, angle and torsion part.}
  \label{fig:Results}
\end{figure}

We tested UFF based BAML using three previously established data sets:
DFT energies and properties of $\sim$7k organic molecules stored in the QM7b data set~\cite{njp}, 
G4MP2 energies and DFT properties for 6k constitutional isomers of C$_7$H$_{10}$O$_2$, 
and DFT energies and properties for 134k organic molecules QM9 (both published in Ref.~\cite{gdb9}).
Links to all data sets are available at {\tt http://quantum-machine.org}. 
Initial structures for all datasets were drawn from the GDB universe~\cite{gdb17, gdb17_anie}.

Log-log plots of BAML learning curves are shown in Fig.~\ref{fig:Results}(a) and (b) for
the C$_7$H$_{10}$O$_2$ isomers as well as for QM9. 
Mean absolute errors (MAEs) for out of sample predictions of nine properties are shown as a function
of training set size for the BAML model. 
Properties studied include enthalpies and free energies of atomization at room temperature, 
heat-capacity at room temperature $C_V$, zero-point-vibrational-energy (ZPVE), norms of dipole moments $\mu$ and polarizability $\alpha$,
as well as HOMO/LUMO eigenvalues and gap. 
For any given property, we find near identical learning rates among BAML models based on 
bonds ($\mathbf{M}^{\mathrm{B}}$), bonds+angles ($\mathbf{M}^{\mathrm{A}}$), 
or atoms+bonds, angles+torsions ($\mathbf{M}^{\mathrm{T}}$), respectively. 
The learning off-set $a$, however, decreases systematically as target similarity to energy
is being increased through the addition 
of higher order contributions---for all properties, and for both data sets. 
We note that BAML can reach chemical accuracy (MAE $\sim$1 kcal/mol with respect to reference) for atomization energies
of the isomers of C$_7$H$_{10}$O$_2$ after training on only 5k example molecules, and MAE $\sim$2.4 kcal/mol for 134k organic molecules in QM9
after training on 10k example molecules. 
To the best of our knowledge, such predictive power has not yet been reached by any other ML model. 
This observation confirms the expectations raised based on the aforementioned arguments. 

In Fig.~\ref{fig:Results}(c)  individual contributions to the atomization energy are on display, 
resulting from  bonds, angles, and torsion representations.  
For illustration, we have selected oultiers, i.e.~three constitutional isomers of C$_7$H$_{10}$O$_2$ for which the out-of-sample prediction error is maximal. 
In all three cases, these molecules experience high internal strain through few membered or joint hetero cycles.
As such, it is reassuring to observer substantial lowering of the error occurs as soon as the representation
accounts explicitly for angular and torsional degrees of freedom. 
Fig.~\ref{fig:Results}(d) indicates averaged changes obtained for the entire constitutional
isomer testing set due to addition of higher
order terms to the representation. 
More specifically, addition of bonds to dressed atoms contributes $\sim$ 15 kcal/mol;
augmenting the representation by angular degrees of freedom reduces this number to $\sim$ 2 kcal/mol;
while further addition of torsional degrees of freedom improves results by mereley $\sim$ 0.5 kcal/mol.
Note that the last change might be small on average, however, for some molecules it
can be consequential if high accuracy shall be achieved, e.g.~in the case of outliers shown in Fig.~\ref{fig:Results}(c).
Overall, it is encouranging to see that these contributions decrease systematically. 
These results suggest that ML models of energies converge rapidly in the many-body expansion of representations.
This finding indicates that it should be possible to construct local ML models
which scale linearly with system size. 

Having established uniqueness and target similarity criteria, we now discuss the choice to represent molecules in terms of bags of interatomic many-body energy expansion. 
This choice is not evident, many representations used in the literature rely on the use of additional properties, such as HOMO/LUMO eigenvalues, or atomic radii and spectra. 
For two reasons we believe an energy based representation to be advantageous: 
First, energy is the very observable associated to the Hamiltonian which defines the system:
The potential energy surface of a given molecular electronic spin-state is equally and uniquely
representative for the system: Two different systems will always differ in their potential energy surface.
Secondly, energy is well understood and there is a large choice of decent energy models, such as UFF, 
which can be used for constructing good representations. 
It is not easy to construct good models of other properties which at the same time also meet the uniqueness criterion. For example, for comparison 
we constructed a molecular representation which reaches high target similarity with another property, polarizability. 
Instead of using bags of interatomic energy contributions we have used a bag of atomic polarizabilities. 
Atomic polarizabilities can easily be obtained from Cartesian coordinates of a molecule, i.e.~without
electronic structure calculations, through the use of Voronoi polyhedra. \cite{OAvL_MBvdW_2}
Unfortunately, this representation violates the uniqueness criterion.
While ($\mathbf{M}^{\mathrm{P}}$) is a decent model of molecular polarizability it is not unique:
Any other molecule which happens to have the same set of atomic volumes, 
irrespective of differences in geometry, will result in the same representation.
Learning curves obtained for $\mathbf{M}^{\mathrm{P}}$ based ML models are shown together with BAML in Fig.~\ref{fig:Results}(a)
for all constitutional isomers. 
All hierarchies of BAML models have lower learning curves for all properties 
except ZPVE and $C_V$ for which the bond based BAML models perform slightly worse. 
In the case of the latter, and for very small unconverged training set sizes, 
the polarizability ML model is even better than any BAML model, 
however, as training set size grows the lack of uniqueness
kicks in with a more shallow learning rate leading to worse performance.
This observation also underscores the necessity to take the convergence behavior of ML models into consideration:
Learning behavior can differ in $a$ {\em and} $b$. 
We note the tendency of the polarizability based ML model towards a smaller slope ($C_V$, ZPVE, $\mu$, $\alpha$), 
indicating the expected lack of uniqueness. 
Surprisingly, even for the target property polarizability $\mathbf{M}^{\mathrm{P}}$ based ML models 
show worse performance than any BAML model.
These numerical results support the idea that Hamiltonian/$\Psi$/Potential Energy Surface 
are  ``special'' in that all other molecular properties can be derived from them, in direct
analogy to the wavefunction $\Psi$, necessary to calculate the corresponding expectation values.

Finally, to place the BAML performance into perspective, 
we have also compared the out-of-sample errors for its most accurate variant to literature results obtained using 
alternative ML models and representations applied to the same molecular data set, QM7b~\cite{njp}.
Tab.~\ref{qm7b_comparison} displays MAEs and RMSE of the ${\bf M}^{\rm T}$ based BAML model trained on 5k molecules, 
as well as London Matrix (LM), Coulomb Matrix (CM), BoB (which is a bag of Coulomb matrix elements), 
and bag of London (BoL) matrix elements. 
In particular, we compare to results from this work with SOAP~\cite{SOAP}, and randomized neural network
based CM model~\cite{njp}. 
SOAP represents a recently introduced sophisticated convolution of kernel, metric and representation. 
We note that by no means this represents a comprehensive assessment, it would have been 
preferable to compare learning curves, such as in Fig.~\ref{fig:Results} (a) and (b). 
However, the implementation of SOAP or neural network models can be complex and is beyond the scope of this study.
BAML yields a MAE for atomization energy of only $\sim$1.15 kcal/mol, only slightly worse than SOAP's $\sim$0.92 kcal/mol. 
We consider such small differences to be negligible for all intents and purposes: Differences between 
QM codes due to use of different compilers, libraries, or optimizers can reach similar scale. 
We also note, however,  in Tab.~\ref{qm7b_comparison} that BAML has a considerably larger RMSE (2.5 kcal/mol) for the atomization energy than SOAP (1.6 kcal/mol). 
Similar observations hold for the polarizability. 
For HOMO/LUMO eigenvalues, ionization potential, electron affinity, and the intensity of the most intense peak, 
BAML yields lowest MAE and lowest RMSE. 
BAML also has the lowest MAE for predicting the first excitation energy (together with the randomized neural network based Coulomb matrix model). The lowest RMSE for this property, however, is obtained using a BOB based ML model. 
The excitation energy of the most intense peak in the model is predicted with the lowest MAE and RMSE when using 
the randomized neural network based Coulomb matrix model. BAML is second for MAE, and third for RMSE (after BoB). 
To further illustrate the effect of target similarity, we also report BoL vs. BoB and LM vs. CM based results.
For most properties, not only the energy, 
the corresponding London variant outperforms the Coulomb element based ML models. 
This finding supports the observations made above that (a) London is more more similar to 
atomization energy than Coulomb, and (b) the more similar the representation to energy, 
the more transferable and applicable it is for other properties.

In conclusion, we have presented arguments and numerical evidence in support of the 
notion that $a$ and $b \log{N}$ in learning curves are influenced, if not determined, 
by the employed representation's target similarity and uniqueness, respectively. 
For molecules, defined by their Hamiltonian which produces their wavefunction which produces the observables, 
BAML models---based on universal force-field parameters and functions for bags of bonds, angles, and torsions---exhibit uniqueness as well as considerable similarity to energy. 
As a result, BAML performs universally well for the modeling of {\em all} simple scalar global 
quantum mechanical observables.
Addition of higher-order contributions in the form of bonds, angles, and torsional degrees of freedom 
enables the systematic lowering of learning curve off-set $a$, resulting in 
BAML models with  unprecedented accuracy, transferability, and speed. 
Finally, we would like to note that it is possible to define a Pareto front 
based on the established learning curves, for which the logarithm of the error
decays as $a - b \log{N}$, and the presented uniqueness and 
target similarity criteria. 
The front is spanned between error and chemical space. 
It negotiates the optimal trade-off between number of training instances 
(or CPU budget in the case of quantum chemistry calculations) and permissible error. 
As such, it is akin to basis set or electron correlation convergence plots, where
for a given computational budget the maximal accuracy can be dialed in, or where, 
conversely, for a given desired accuracy, the necessary computational budget can 
be estimated. 
This latter aspect might be relevant to the automatized generation of QM derived
property models with predefined uncertainty and transferability.



\begin{acknowledgements}
O.A.v.L. acknowledges funding from the Swiss National Science foundation (No.~PP00P2\_138932, 310030\_160067).
This research was partly supported by the NCCR MARVEL, funded by the Swiss National Science Foundation.
Calculations were performed at sciCORE (http://scicore.unibas.ch/) scientific computing core facility at University of Basel. 
\end{acknowledgements}


\bibliography{MBP}

\begin{thebibliography}{32}
\bibitem [{SID()}]{SIDENOTE}%
  \BibitemOpen
  \href@noop {} {}\bibinfo {note} {This simple example also underscores the
  qualitative importance of many-body contributions in interatomic energy
  decompositions which play a role even in effects as weak as London dispersion
  forces---in addition to the already established quantitative role they play
  in nature \cite{OAvL_MBvdW_1, OAvL_MBvdW_2}.}\BibitemShut {Stop}%
\end{thebibliography}

\end{document}